\documentclass[a4paper,12pt]{article}
\pdfoutput=1
\usepackage{epsfig}
\usepackage{amssymb}
\usepackage{amsfonts}
\usepackage{amsmath}
\usepackage{euscript}
\usepackage{verbatim}
\usepackage{latexsym}
\usepackage{graphicx}
\usepackage{caption}
\usepackage{float}
\usepackage{subcaption}

\newif\ifdtup

\catcode`\@=11

\@addtoreset{equation}{section}

\def\@normalsize{\@setsize\normalsize{15pt}\xiipt\@xiipt
\abovedisplayskip 14pt plus3pt minus3pt%
\belowdisplayskip \abovedisplayskip
\abovedisplayshortskip \z@ plus3pt%
\belowdisplayshortskip 7pt plus3.5pt minus0pt}

\def\small{\@setsize\small{13.6pt}\xipt\@xipt
\abovedisplayskip 13pt plus3pt minus3pt%
\belowdisplayskip \abovedisplayskip
\abovedisplayshortskip \z@ plus3pt%
\belowdisplayshortskip 7pt plus3.5pt minus0pt
\def\@listi{\parsep 4.5pt plus 2pt minus 1pt
     \itemsep \parsep
     \topsep 9pt plus 3pt minus 3pt}}

\relax

\catcode`@=12

\catcode`\@=11

\def\section{\@startsection{section}{1}{\z@}{3.5ex plus 1ex minus
   .2ex}{2.3ex plus .2ex}{\large\bf}}

\def\SymBoxes#1#2#3#4{\newdimen\un@t \un@t#3%
\raisebox{#1}{\rule{#2\un@t}{#4}\hskip-#2\un@t
\@tempdimb\un@t \advance\@tempdimb by-#4\@tempcntb#2\relax%
\@whilenum{\@tempcntb>0}\do{
\rule{#4}{\un@t}\hskip\@tempdimb \advance\@tempcntb by\m@ne}%
\hskip-#2\un@t \rule[\un@t]{#2\un@t}{#4}%
\rule[\un@t]{#4}{#4}\hskip-#4
\rule{#4}{\un@t}}\hskip-#4}                

\begin{document}

\newcommand{\beq}{\begin{equation}}
\newcommand{\eeq}{\end{equation}}
\newcommand{\bea}{\begin{eqnarray}}
\newcommand{\eea}{\end{eqnarray}}
\newcommand{\beas}{\begin{eqnarray*}}
\newcommand{\eeaas}{\end{eqnarray*}}
\newcommand{\defi}{\stackrel{\rm def}{=}}
\newcommand{\non}{\nonumber}
\newcommand{\bquo}{\begin{quote}}
\newcommand{\enqu}{\end{quote}}
\renewcommand{\(}{\begin{equation}}
\renewcommand{\)}{\end{equation}}
\def \eqn#1#2{\begin{equation}#2\label{#1}\end{equation}}

\def\e{\epsilon}
\def\IZ{{\mathbb Z}}
\def\IR{{\mathbb R}}
\def\IC{{\mathbb C}}
\def\IQ{{\mathbb Q}}
\def\de{\partial}
\def\Tr{ \hbox{\rm Tr}}
\def\H{ \hbox{\rm H}}
\def\HE{ \hbox{$\rm H^{even}$}}
\def\HO{ \hbox{$\rm H^{odd}$}}
\def\K{ \hbox{\rm K}}
\def\Im{ \hbox{\rm Im}}
\def\Ker{ \hbox{\rm Ker}}
\def\const{\hbox {\rm const.}}
\def\o{\over}
\def\im{\hbox{\rm Im}}
\def\re{\hbox{\rm Re}}
\def\bra{\langle}\def\ket{\rangle}
\def\Arg{\hbox {\rm Arg}}
\def\Re{\hbox {\rm Re}}
\def\Im{\hbox {\rm Im}}
\def\exo{\hbox {\rm exp}}
\def\diag{\hbox{\rm diag}}
\def\longvert{{\rule[-2mm]{0.1mm}{7mm}}\,}
\def\a{\alpha}
\def\dag{{}^{\dagger}}
\def\tq{{\widetilde q}}
\def\p{{}^{\prime}}
\def\W{W}
\def\N{{\cal N}}
\def\hsp{,\hspace{.7cm}}

\def\br{\nonumber}
\def\IZ{{\mathbb Z}}
\def\IR{{\mathbb R}}
\def\IC{{\mathbb C}}
\def\IQ{{\mathbb Q}}
\def\IP{{\mathbb P}}
\def \eqn#1#2{\begin{equation}#2\label{#1}\end{equation}}

\newcommand{\C}{\ensuremath{\mathbb C}}
\newcommand{\Z}{\ensuremath{\mathbb Z}}
\newcommand{\R}{\ensuremath{\mathbb R}}
\newcommand{\rp}{\ensuremath{\mathbb {RP}}}
\newcommand{\cp}{\ensuremath{\mathbb {CP}}}
\newcommand{\vac}{\ensuremath{|0\rangle}}
\newcommand{\vact}{\ensuremath{|00\rangle}                    }
\newcommand{\oc}{\ensuremath{\overline{c}}}
\newcommand{\psizero}{\psi_{0}}
\newcommand{\phizero}{\phi_{0}}
\newcommand{\hzero}{h_{0}}
\newcommand{\psiin}{\psi_{\rh}}
\newcommand{\phiin}{\phi_{\rh}}
\newcommand{\hin}{h_{\rh}}
\newcommand{\rh}{r_{h}}
\newcommand{\rb}{r_{b}}
\newcommand{\psibnd}{\psi_{0}^{b}}
\newcommand{\psibndp}{\psi_{1}^{b}}
\newcommand{\phibnd}{\phi_{0}^{b}}
\newcommand{\phibndp}{\phi_{1}^{b}}
\newcommand{\gbnd}{g_{0}^{b}}
\newcommand{\hbnd}{h_{0}^{b}}
\newcommand{\zh}{z_{h}}
\newcommand{\zb}{z_{b}}
\newcommand{\man}{\mathcal{M}}
\newcommand{\hbr}{\bar{h}}
\newcommand{\tbr}{\bar{t}}

\begin{titlepage}
\begin{flushright}
CHEP XXXXX
\end{flushright}
\bigskip
\def\thefootnote{\fnsymbol{footnote}}

\begin{center}
{\Large
{\bf de Sitter, $\alpha'$-Corrections \&  Duality Invariant 
\\ 
Cosmology \\
}
}
\end{center}

\bigskip
\begin{center}
Chethan KRISHNAN$^a$\footnote{\texttt{chethan.krishnan@gmail.com}}
\vspace{0.1in}

\end{center}

\renewcommand{\thefootnote}{\arabic{footnote}}

\begin{center}

$^a$ {Center for High Energy Physics,\\
Indian Institute of Science, Bangalore 560012, India}\\

\end{center}

\noindent
\begin{center} {\bf Abstract} \end{center}
Demanding $O(d,d)$-duality covariance, Hohm and Zwiebach have written down the action for the most general cosmology involving the metric, $b$-field and dilaton, to all orders in $\alpha'$ in the string frame. Remarkably, for an FRW metric-dilaton ansatz the equations of motion turn out to be quite simple, except for the presence of an unknown function of a single variable. If this unknown function satisfies some simple  properties, it allows de Sitter solutions in the string frame. In this note, we write down the Einstein frame analogues of these equations, and make some observations that make the system tractable. Perhaps surprisingly, we find that a necessary condition for de Sitter solutions to exist is that the unknown function must satisfy a certain second order non-linear ODE. The solutions of the ODE do not have a simple power series expansion compatible with the leading supergravity expectation. We discuss  possible interpretations of this fact. After emphasizing that all (potential) string and Einstein frame de Sitter solutions have a running dilaton, we write down the most general cosmologies with a constant dilaton in string/Einstein frame: these have power law scale factors. 




\vspace{1.6 cm}
\vfill

\end{titlepage}

\setcounter{footnote}{0}

\section{Introduction}

Even though known constructions of de Sitter vacua in string theory \cite{KKLT, LVS} rely on finite quantum corrections\footnote{And are also controversial \cite{Willy, Westphal, Danielsson, Sethi, Trivedi}. See \cite{Gutperle, Ulf, Jon} for some alternative approaches to dS-like physics in string inspired set ups.}, there is no known No-Go Theorem that definitively dictates that tree level string theory can{\em not} have de Sitter vacua \cite{Silverstein}. 

For various subcategories of tree level constructions however, there do exist claims of varying confidence, that forbid de Sitter vacua. The de Sitter swampland conjectures\footnote{See \cite{Vafa1} for the initial proposal, \cite{Garg1, Vafa2} for the potentially viable version, and \cite{Andriot2, Garg2} for further refinements.}, when interpreted as statements about tree level string theory with singular sources, are candidates for such No-Go theorems. Some of these bounds could very well be  universal statements potentially provable in reliable tree level set ups\footnote{By reliable, we mean that there are no unfixed moduli at positive vacuum energy, and therefore  quantum corrections can be consistently ignored. See \cite{Dibitetto} for an example that violates the dS swampland bounds by violating this condition. Ideally, for a reliable solution, one would also like parametric control.}. But at the moment all these bounds are conjectural, and do not qualify as No-Go {\em theorems}. The evidence for them mostly comes from supergravity with singular sources. 


In the absence of a watertight  No-Go theorem, one might seek hints for the (non-)existence of tree level de Sitter vacua from a different angle. One such angle is the systematic inclusion of $\alpha'$-corrections to the  low energy effective action of string theory\footnote{For efforts to explicitly construct de Sitter in string theory using $\alpha'$-effects and/or higher derivative corrections, see eg. \cite{Mat, Ulf2}.}. This gives a somewhat complementary approach, because it goes beyond supergravity by incorporating higher derivative corrections. If one can determine the $\alpha'$-corrections, then one has a new context for discussing the existence of de Sitter vacua. However, explicit calculation of these corrections has turned out to be difficult beyond low orders. Since truncating a theory at a finite number of orders often leads to unphysical features, it is difficult to make a reliable statement about the (non-)existence of de Sitter without taking all $\alpha'$-corrections into account \cite{Hohm1,   Hohm2}\footnote{This means that if a de Sitter exists, it will have to be non-perturbative \cite{Hohm1, Hohm2} in $\alpha'$.}. 

A way out is to drop the goal of direct calculation of the $\alpha'$ corrections in string theory, and instead use T-duality covariance as the tool to determine them. This strategy leads to $O(d,d)$-duality covariant field theory \cite{Aldazabal:2013sca, Hohm:2013jaa}. String theory effective actions will then be points in the theory space of such duality covariant theories. Even this turns out to be too much to ask however: complete all order corrections in generic situations is not known. But Hohm and Zwiebach have recently shown \cite{Hohm1, Hohm2} that in cosmological settings such corrections can be determined to all orders in $\alpha'$ for the metric, $b$-field and dilaton. This is the context of the present paper.

One conclusion of \cite{Hohm1, Hohm2} is that when one restricts attention further to an FRW ansatz, it turns out that the system drastically simplifies: it becomes a system of ODEs for two fields\footnote{Loosely, the scale factor and the dilaton.}, controlled by one unknown function of one variable, see \eqref{F}. In fact, it becomes possible to argue that if the unknown function has some simple properties, the FRW ansatz admits a de Sitter solution. This is interesting, in light of the fact that tree level constructions in string theory, by which one usually means\footnote{Somewhat imprecisely!}  supergravity with singular sources, have so far failed to construct a stable de Sitter vacuum. Perhaps the message is that one should include $\alpha'$ corrections systematically, if one wants to find a de Sitter vacuum?

There are at least three causes for pause however. First one is that the solutions found in \cite{Hohm1, Hohm2} have a non-constant dilaton. A constant $O(d,d)$-dilaton, as found in \cite{Hohm1, Hohm2}, together with a time-dependent scale factor necessarily implies a time-dependent physical dilaton\footnote{One curious fact is that the time-dependence leads to a linear dilaton. It will be interesting to see if this can be put to some use.}. To emphasize this, we will write down the most general solution with a constant dilaton in a later section, and it turns out that they are at best power law cosmologies.

The second is that it is unclear whether the solutions in \cite{Hohm1, Hohm2} are stable. As known for some time (see eg. \cite{Wrase}), tachyonic de Sitter solutions are reasonably straightforward to come by: it is stable de Sitter vacua, that are hard to obtain \cite{Garg1, Vafa2}. However, the question of stability of the solutions in \cite{Hohm1, Hohm2} is an issue that is probably difficult to make progress in, without knowledge of explicit forms for the unknown function in \eqref{F}. So we will not undertake it here\footnote{But perhaps the fact that we are able to make statements about de Sitter in Einstein frame is also somewhat surprising in hindsight. So it is conceivable that we can make progress on the stability question as well.}.

The third issue is that the dS solutions found in \cite{Hohm1, Hohm2} are string frame solutions. This is the launching point for the present paper: we will primarily be concerned with the Einstein frame, largely because the fields in the string frame do not have canonical kinetic terms\footnote{Some authors seem more relaxed about the choice of frame (see discussions in eg. \cite{Angus}), but we have an interest in loose qualitative comparisons with conventional tree level string constructions which are in the Einstein frame, so we will stick to this. But note that in constant dilaton solutions, the difference in frame choice is mostly cosmetic.}. Our goal will be to investigate the possibility of de Sitter solutions. We will use some tricks to make the system tractable, and find that perhaps surprisingly, a necessary condition for the existence of de Sitter solutions is that the unknown function satisfy a certain non-linear ordinary differential equation. We will find that this ODE does not have a Frobenius series solution that matches the leading supergravity expectation, so we make some comments about whether it is possible for the system to have de Sitter solutions. 


Despite the somewhat defeatist nature of the results we find in this paper, we feel that the approach of \cite{Hohm1, Hohm2} is an intriguing one. We will say more about this, in a concluding section. Indeed, it would be very interesting to incorporate more fields and investigate the all order equations, to address the possibility of a stable de Sitter vacuum  in the Einstein frame with at least a quasi-constant dilaton. One thing we find worthy of note in our results is the very striking differences between the structure of dS solutions in the string frame and the Einstein frame. We were not expecting to find that we could extract a stringent condition (a 2nd order ODE) on the unknown function by demanding the existence of a de Sitter solution. Remarkably, we found a local (in the space of arguments of the unknwon function) condition on the function. Note that in the string frame, the conditions on the function were global statements in the space of arguments about the existence of certain zeros.

\section{Double Field Theory, $O(d,d)$ Duality, String Frame Cosmology}

We will start with a quick review of some of the results in \cite{Hohm1, Hohm2} to set up notation. Any self-containedness of the presentation should be viewed as incidental.

The idea is to write down the most general corrections to the  equations of motion by demanding that they respect $O(d,d,\IR)$ duality symmetry. The starting point is the $D= d+1$ 
dimensional two-derivative string frame action of closed string theory for the massless NSNS fields
\bea
\label{stringaction}
I \ = \   \int d^Dx  \sqrt{-g}\,  e^{-2\phi} \Bigl( R +  4 (\partial \phi)^2 -\frac{1}{12} H^2 +\cdots\Bigr)
\eea
where $H_{\mu\nu\rho}  =  3  \partial_{[\mu} b_{\nu\rho]}$.

Computing $\alpha'$-correction for general backgrounds is too difficult at present, even after demanding duality invariance. But if one restricts to cosmological backgrounds (ie., the only dependence is on time $t$), it turns out that we can determine the all-order corrections \cite{Hohm1, Hohm2}. To explain this we set $x^\mu   =  (t, x^i)$, $i  =  1,\ldots, d$ and demand all spatial derivatives be zero. 
For the metric, $b$-field, and dilaton we set:
\bea\label{ansatz}
g_{\mu\nu}\ = \ \begin{pmatrix} -n^2  (t) & 0 \\ 0 & g_{ij} (t) \end{pmatrix},\ \ \ \
b_{\mu\nu} \ = \ \begin{pmatrix} 0 & 0 \\ 0 & b_{ij} (t) \end{pmatrix}, \ \ \ \  
\phi \ = \ \phi (t) . 
\eea
Reducing the two-derivative terms in (\ref{stringaction}) to one dimension one obtains \cite{Veneziano:1991ek}:  
\bea
\label{reducedaction} 
I_0 \ = \ 
  \int d t  \, e^{-\Phi}\,  {1\over n} \,\Bigl(  -\dot \Phi^{\,2} -   \frac{1}{8}
  {\rm tr} \big(\dot{\cal S}^2\big) \,  \Bigr). 
\eea
Here the generalized metric  ${\cal S}$ is a $2d\times 2d$ matrix
\bea
\label{genmetric}
{\cal S} \ \equiv \ 
\begin{pmatrix}  bg^{-1} & g - b g^{-1} b \\[0.7ex]
 g^{-1} & - g^{-1} b 
\end{pmatrix} ,
\eea
that is $O(d,d,\IR)$ valued and satisfies ${\cal S}^2  =  {\bf 1}$. The new dilaton $\Phi$ is fixed by    
 \bea\label{Odd-dilaton}
  e^{-\Phi} \ \equiv \ \sqrt{\det g_{ij}} \,e^{-2\phi},
 \eea 
and is $O(d,d,\IR)$-invariant. The reduced action has manifest $O(d,d,\IR)$ invariance under $\Lambda$:  
\bea\label{duality}
{\cal S}\ \rightarrow \ {\cal S}' \ = \  \Lambda \,{\cal S} \,\Lambda^{-1}\;, \qquad  \hbox{with} \qquad  
\Lambda \,\eta\, \Lambda^t \ = \ \eta\;, \qquad 
\eta \ = \ \begin{pmatrix}  0 & {\bf 1} \\[0.7ex]
 {\bf 1} & 0 
\end{pmatrix},  
\eea
with $\eta$ being the $O(d,d,\mathbb{R})$ invariant metric.

In \cite{Hohm1, Hohm2} the most general $\alpha'$-corrections to $I_0$ above were written down building on the work of \cite{Veneziano:1991ek,Meissner:1996sa,added}. The most general form can be found in eqn. (1.3) of \cite{Hohm1}, and we will not reproduce it here. The general form contains undetermined coefficients, which we will collectively denote by $c_k$ here, and $c_1$ is fixed by the leading supergravity result that we have written down above. 
The $c_k$ beyond $c_1$ (as well as the higher trace coefficients) are not fixed by duality invariance. For any given string theory, these coefficients can in principle be determined and this determines the complete 
$\alpha'$ corrected dynamics of the cosmological background in that string theory. There are three ODEs that arise as the equations of motion for $\Phi$, ${\cal S}$ and $n$ from this action.  Since there are infinite number of coefficients which we do not know the form of, this is still quite a non-trivial challenge to make progress in. But further simplifications happen when we restrict ourselves further to an FRW ansatz, which is sufficient to discuss de Sitter.


To see this, we specialize to an FRW ansatz,
 \bea
  ds^2  =  -dt^2  +  a^2(t)\, d{\bf x}^2, 
 \eea
where $d{\bf x}^2$ is the flat Euclidean spatial piece
and $a(t)$ is the scale factor.  With the $b$-field set to zero, the generalized metric becomes
 \bea\label{S-FRW}
  {\cal S}(t) =  \begin{pmatrix}  0 & a^2(t) \\[0.7ex]
 a^{-2}(t) & 0 
\end{pmatrix}\;.
 \eea 
It turns out that the generalized Friedmann equations that follow from the all order corrected action above can now be written as\footnote{The first of these equations can be viewed as a Bianchi identity and is therefore redundant, so in what follows we will only work with the last two equations. Our results put constraints on the function $F$, and it is straightforward to check that for {\em any} choice of $F$, the first equation can be obtained from the other two.} \cite{Hohm1, Hohm2}
 \bea\label{stringframeEOM}
  \ddot\Phi + \tfrac{1}{2}H f(H) \ &= \ 0\;, \label{stringframeEOM1}\\ 
  \frac{{ d}}{{ d}t}\big(e^{-\Phi} f(H) \big) \ &= \ 0
   \;, \label{stringframeEOM2}\\
   \dot{\Phi}^2+\, g(H) \ &= \ 0\;,\label{stringframeEOM3}
 \eea  
with $H(t)$ the Hubble ``constant"  $H(t) = \frac{\dot{a}(t)}{a(t)}$, and $f$ and $g$ can be defined via a single (even!) function $F(H)$ where
 \bea \label{F}
  F(H) \equiv 4d\sum_{k=1}^{\infty}(-\alpha')^{k-1}\, 2^{2k-1}\,c_k\, H^{2k} \ = \ -d\, H^2 +\cdots \;, 
 \eea
Through $F$, $f(H)$ and $g(H)$ are defined as\footnote{Note that $g$ is very closely related to the Legendre transform of $F$, except that if $g$ were a ``true" Legendre transform it would be viewed as a function of $F'$  and not $H$.}
\bea
  f(H)  =  F'(H)\;, \qquad g(H) = HF'(H)-F(H) \label{dfn}
 \eea 
In extracting the leading order expression, we have used $c_1 = -\frac{1}{8}$. This implies that
  \bea
  f(H) &=  d\,  \sum_{k=1}^{\infty}(-\alpha')^{k-1}\, 2^{2(k+1)} \, k\, c_k \,H^{2k-1}  \,\,  = \ -2 d\,  H + \cdots ,\label{fexpansion} \\
  g(H)  &= \ d \sum_{k=1}^{\infty}(-\alpha')^{k-1}2^{2k+1}(2k-1) c_k\, H^{2k} = - d  \, H^2 + \cdots\label{g-expansion}
 \eea 
and also that they satisfy
 \bea\label{interesting}
  g'(H) \ = \ H f'(H).
 \eea 
Using the fact that $g(0)=0$, we can rewrite this also as 
\bea\label{g}
  g(H) \ = \ Hf(H)-\int_0^H f(H'){d}H'. 
 \eea
 
One observation made in \cite{Hohm1, Hohm2} was that these equations admit de Sitter solutions if $f$ and $g$ have a common zero at a non-zero value of $H=H_0$. It is easy to convince oneself that this is indeed possible to arrange, with appropriate choices of $F$, and an explicit example was given in \cite{Hohm1, Hohm2}. Quite generally, it can be seen that any even function $F(H)$ that behaves as $\sim -H^2$ for $H \to 0$, and also has non-vanishing arguments where $F$ and $F'$ simultaneously vanish will lead to de Sitter solutions. It is also worth noting that the $O(d,d)$-dilaton $\Phi$ is constant in these solutions\footnote{But note that this implies that the dilaton $\phi$ itself is in fact, time dependent. We think this is worth a note.}.

Of course, whether such an $O(d,d)$-field theory function $F$ exists in any specific  string theory is a question that was not addressed in \cite{Hohm1, Hohm2}. We will also not worry about this issue: 
We will take a more ``bottom-up" approach and try to see what functions $F$ can possibly admit de Sitter solutions.

\section{Hohm-Zwiebach-FRW Equations in the Einstein Frame}

As stated earlier, our first goal in this paper is to write down the Einstein frame analogues of \eqref{stringframeEOM}. we want to get to Einstein frame metrics that are of the FRW form
\bea
ds^2_{E}=-dT^2+a_E(T)^2 dx_i^2 
\eea
by starting with the string frame metric
$ ds^2_{S}=-dt^2=a(t)^2 dx_i^2$ where $a(t)$ is the string frame scale factor we discussed in the last section, $a_E$ is the Einstein frame scale factor, and we have noted that the time coordinates allow a possible reparametrization by calling the Einstein frame time $T$ instead of $t$. Using the relation between string frame and Einstein frame $g_{\mu\nu}^E=e^{-\frac{4 \phi}{d-1}}\ g_{\mu \nu}^S$, we immediately get the two key relations
\bea
a_E(T) = a(t) \ e^{-2\phi(t)/(d-1)}, \ \ \ 
\frac{dT}{dt} = e^{-2\phi(t)/(d-1)}.
\eea
Using these we can also write the $O(d,d)$-dilaton in a few useful forms:
\bea
e^{-\Phi}=a^d \, e^{-2 \phi}= a_E^d \, e^{2 \phi/(d-1)} = a_E^{d-1}\, a.
\eea
The independent Hohm-Zwiebach equations \eqref{stringframeEOM2}, \eqref{stringframeEOM3} can be viewed as two equations for the two independent fields $a(t)$ and $\Phi(t)$. But we will find it convenient to work with a change of variables, when we are in the Einstein frame. There are a few choices for the two variables that one can use to write down the equations of motion in the Einstein frame: one of the two is naturally $a_E(T)$, but should the other variable be $\Phi$ or $\phi$, or something else altogether? We will find it convenient to choose the two variables to be $a_E(T)$ and $a(T)$, the latter is the scale factor in the string frame but written in terms of $T$ \footnote{In practice, this just means that we are working with $a_E(T)$ and $a(T) \equiv a_E(T) \ e^{2\phi(T)/(d-1)}$ as the two independent variables. In particular, we are {\em not} making any new assumptions. The dilaton is implicitly present, and we can change variables back to it, if we wish. Note also that any function $f(t)$ in the string frame time defines an $f(T)$ via $f(t(T))$ in the Einstein frame time, which is what the notation captures. }. Writing \eqref{stringframeEOM2} and \eqref{stringframeEOM3} in terms of these variables leads to the two equations
\bea
 f\Big(\frac{a_E}{a^2}\frac{da}{dT}\Big)& =& \frac{k}{a_E^{d-1}a},  \\
g\Big(\frac{a_E}{a^2}\frac{da}{dT}\Big) &=&- \frac{a_E^2}{a^2}\left(\frac{(d-1)}{a_E}\frac{da_E}{dT}+\frac{1}{a}\frac{da}{dT} \right)^2
\eea
where $k$ is an arbitrary integration constant\footnote{The integration constant simply arises from integrating the total differential in \eqref{stringframeEOM2}.}. Because of the presence of the arbitrary function $F$ that controls $f$ and $g$, it is not immediately clear how we can proceed from here. 

It is clear however from the equations that the case $d=1$ is going to be different, so let us tackle that first. The second equation then simplifies qualitatively, and implies that the unknown function $F(x)$ is determined via $g(x)=-x^2$. Note that this expression is precisely the $(\alpha')^0$ expression for $g$ coming from supergravity \eqref{g-expansion} when $d=1$. If we demand that the Einstein frame metric takes the de Sitter form, with $H_E$ constant, one can explicitly solve for $a$ using the first equation and one finds $\ln a(T)/a(0) \sim (1- e^{-H_ET})$. Note however that the dilaton is degenerate in $d=1$ dimension, so we will not pursue this case further. 

However, the structure of the $d=1$ case suggests a potential way to proceed: we should find a way to constrain the function $F(x)$ (or equivalently, $f(x)$ and $g(x)$). It is perhaps not immediately clear how this can be accomplished, but a useful first step\footnote{To motivate some of the tricks below, it is helpful to introduce the inverse function of $f$. But it is crucial that we are able to make the argument without relying on the inverse function to avoid getting caught up in well-definedness and uniqueness issues.} is to write the second equation of motion in the form
\bea
-\frac{a_E}{a^2}\frac{da}{dT}\pm \sqrt{-g\Big(\frac{a_E}{a^2}\frac{da}{dT}\Big)}=\frac{a_E}{a}(d-1) H_E
\eea
 and then use the first equation of motion to re-write the right hand side. Note that we have introduced the notation $H_E\equiv \frac{1}{a_E}\frac{d a_E}{dT}$, but we are not (yet) demanding that $H_E$ be constant. After some simple manipulations we get
\bea
\frac{-\frac{a_E}{a^2}\frac{da}{dT}\pm \sqrt{-g\Big(\frac{a_E}{a^2}\frac{da}{dT}\Big)}}{f\Big(\frac{a_E}{a^2}\frac{da}{dT}\Big)}=\frac{a_E^d (d-1) H_E}{k} .
\eea
This is a key relation, and if we demand that the system admit an Einstein frame de Sitter solution for $a_E(T)$, it lets us (in principle) solve the system completely! This happens via a determination of the form of $F(H)$ through a second order non-linear ODE for $F(H)$. We do not have a deep understanding of why this has become possible, but it is easy enough to demonstrate, as we do below.

Note that an Einstein frame de Sitter solution has $a_E(T)=a_E(0)e^{H_E T}$, with $H_E$ constant. Therefore we can write the above relation as
\bea
\alpha(T)=\frac{-x\pm \sqrt{-g(x)}}{\beta f(x)} \label{key}
\eea 
where
\bea
x(T)\ \equiv \frac{a_E}{a^2}\frac{da}{dT}, \ \ \alpha(T)\equiv \frac{a_E(T)^d}{k}, \ \ \beta\equiv (d-1)H_E.
\eea
A key point is that the right hand side only depends on $T$ implicitly through $x$, whereas the left hand side is an explicit (and known) function of $T$ which has the property that 
\bea
\frac{d \alpha}{dT} =  H_E \, \alpha(T) d
\eea
Therefore by taking a time derivative of \eqref{key} and using the equality between the first and last expressions in 
\bea
\mp\frac{\sqrt{-g(x)}}{a_E(T)^d}=\frac{d}{dT}\left(\frac{1}{a_E^{d-1}a}\right)=\frac{d}{dT}\frac{f(x)}{k}=\frac{f'(x)}{k}\frac{dx}{dT} \label{dxdtsimple}
\eea
to write 
\bea
\frac{dx}{dT}=\mp\frac{1}{\alpha(T)}\frac{\sqrt{-g(x)}}{f'(x)}
\eea
and using \eqref{key} again to write $\alpha(T)$ in terms of functions of $x$, we get a differential equation {\em purely} in  $x$. In \eqref{dxdtsimple}, the first equality is a version of the second EOM, and the second equality is a consequence of the first EOM. 

The final ODE is most conveniently written as
\bea
\pm\frac{d}{d-1}\frac{f'(x)}{\sqrt{-g(x)}}=\frac{d}{dx}\left({\frac{f(x)}{-x\pm\sqrt{-g(x)}}}\right).\label{ODE}
\eea
Together with the definitions \eqref{dfn} of $f$ and $g$ in terms of $F$, this leads to a second order non-linear ODE for $F$. Because it is an ODE, it is trivial for Mathematica to numerically integrate it for various choices of $F$ and $F'$ from some $x=x_0$. However, to qualify as an acceptable solution, it must match in an appropriate sense with the leading supergravity results in \eqref{F}. Note that given such an acceptable $F$, the entire problem is essentially solved, because all the equations are ODEs that are straightforwardly integrated numerically.

\section{What is an Acceptable $F$?}

The solutions for $F$ should match with the leading supergravity expectation in \eqref{F}, but what is an appropriate way to interpret this statement? The all-order $\alpha'$ results of \cite{Hohm1, Hohm2} are perturbative results. This means that they should perhaps not be interpreted as power series expansions with a finite radius of convergence, and they should be viewed as asymptotic expansions in the worldsheet sigma model. Indeed, if we choose to interpret the $\alpha'$ expansion as having a finite radius of convergenc, we will see that the arguments that we present in the next paragraphs rule out the existence of a de Sitter solution in the Einstein frame. But in the spirit of \cite{Hohm1, Hohm2}, we would like to consider the best-case-scenario for the existence of de Sitter, so we will  consider the possibility that the $\alpha'$-expansion is an asymptotic expansion\footnote{But note that if one views the $\alpha'$ expansion as loosely dual to an expansion in the 't Hooft coupling $\lambda$, then it is quite natural for it to have a finite radius of convergence. This is because while the number of Feynman diagrams explodes exponentially, the number of Feynman diagrams at genus zero in the large-$N$ limit does not. If one takes this heuristic argument seriously, the discussion in this paper is sufficient to rule out the existence of de Sitter solutions in the Einstein frame. But while this argument may apply to some quantities (say, certain scattering amplitudes), it is difficult to formulate such an argument for something like an effective action, which is the type of thing we are intersted in. I thank Ashoke Sen for related comments.}.

In an asymptotic expansion, we expect the $k$-th coefficient to grow factorially in $k$. Instead if we demand that $F$ has a Taylor series expansion\footnote{The argument of the functions $f$, $g$, and $F$ we call either $x$ or $H$.} 
\bea
F(H)=F_0 +F_1 H + F_2 H^2 + ...
\eea
around $H=0$, then it can be immediately demonstrated by directly plugging into the ODE \eqref{ODE} that a solution with the correct structure near $H=0$, namely
\bea
F = -d \, H^2 + ...
\eea
does {\em not} exist\footnote{In fact, a stronger statement can be made, namely that the series can not have a Frobenius series type solution (ie., one where $H^w$ multiplies a Taylor series expansion, with any real $w$), if one wants to match with supergravity.}. Note also that if we take the stand that an expansion with a finite radius of convergence around $H=0$ will necessarily {\em not} represent $F$,\footnote{Note that this statement is quite general and does not depend on the frame.} it will mean that the example function (a sinusoid) that was suggested in \cite{Hohm1, Hohm2} as a candidate for constructing de Sitter solutions in string frame, cannot be realistic. Let us emphasize however that the only fact one needs for the existence of a string frame de Sitter in \cite{Hohm1, Hohm2} is that both $F$ and $F'$ vanish at some finite value of the argument. 

These discussions mean that the mismatch between the supergravity expectation and the Taylor expansion is not quite an indication that $F$ cannot exist, if $F$ is instead viewed as an asymptotic expansion.  In order to interpret $F$ as a function while it is only defined as an asymptotic expansion, we have to implicitly adopt {\em some} philosophy about its nature as a series: what the argument above demonstrates is merely that it cannot be an analytic function with a finite radius of convergence around $H=0$. If we assume that the function should incorporate non-perturbative corrections in its series expansion (eg., like a resurgent transeries), then the above arguments will not rule out the existence of an Einstein frame de Sitter.


So what is an appropriate choice for a series expansion for $F$ that incorporates non-pertubative effects? We will not be able to conclusively answer this question in this paper, but let us expand slightly on the above comments.

It is worth noting that one can write the form \eqref{F} as
\bea
F=\frac{1}{\alpha'}F_0(\sqrt{\alpha'} H)
\eea
for some even $F_0$. This means that one can view the small $\alpha'$ expansions as small $H$ expansions. We can make guesses about the form of the non-perturbative corrections, let us consider as an example
\bea
F(H)=\frac{1}{\alpha'}F_0(\sqrt{\alpha'}H)\sim\sum_k c_k\alpha'^k H^{2k}+\exp\Big(-\frac{1}{\sqrt{\alpha'}H} \Big)\Big(\sum_n {c'}_n\alpha'^nH^{2n} \Big).
\eea 
as $\alpha' \rightarrow 0$. Here we have used the fact that only even powers show up in \eqref{F}. The specific form for the non-perturbative effects we have written down is only for illustrating the possibility. It will be interesting to see if {\em some} series that incorporates non-perturbative correction(s) can actually solve the ODE so that the leading perturbative part is consistent with the leading SUGRA result \eqref{F}. It will be very interesting if this turns out to be true: this would mean that non-perturbative $\alpha'$ corrections could at least in principle produce de Sitter solutions\footnote{Note that this is somewhat different from the sense in which the word ``non-perturbative" was used in \cite{Hohm1, Hohm2}; in particular, in the context of their example sinusoid function. There, summing a Taylor expansion around the origin to write down the closed form of the function was also called non-perturbative. Here instead, we are working with asymptotic expansions.}. 

A complementary angle on the problem is that we can try to solve the ODE directly, analytically. Since this is a non-linear ODE, usual Frobenius method does not immediately work: one has to either find clever guesses for successful power series expansions\footnote{To see an example problem where such a guess was successfully made (in a completely different context), see \cite{Daniel}.}, possibly inspired by non-perturbative expectation like above, or try to connect it with some approach like Painlev\'e theory. To do the latter, one has to bring the ODE \eqref{ODE} to the form 
\bea
\frac{d^2y}{dx^2}=G\left(\frac{dy}{dx},y,x\right)
\eea
where $G$ is rational in $y$ and $dy/dx$ and analytic in $x$. It is not immediately clear how this can be done, even though we cannot rule it out a-priori. One further complication that non-linear ODEs bring to the table is that they can have ``movable" singular points, which do not merely depend on the structure of the ODE, but also on the integration constant. Painlev\'e theory applies when this does {\em not} happen, so it remains to be seen whether this is a viable approach.

It is not clear to us whether some physically motivated (worldsheet instantons?) ``ansatz" could work as a candidate series. It will certainly be interesting if some progress along these lines can be made. But since treating the $\alpha'$-expansion as having a finite radius of convergence (like it is done in \cite{Hohm1, Hohm2}), leads to the clean result that an Einstein frame de Sitter cannot exist, that is where we will leave the present discussion.

\section{Constant Dilaton Cosmology}

Before we conclude, let us also note the case of constant dilaton, which is the situation that one typically thinks of as the scenario with the greatest physical relevance. Instead of demanding the scale factor is that of de Sitter, we will now demand that the dilaton is a constant, and see whether there can be any interesting solutions for the scale factor.

We will write the equations of motion in the Einstein frame in the form
\bea
f\left(e^{-2\phi/(d-1)}\Big(\frac{1}{a_E}\frac{da_E}{dT}+\frac{2}{d-1}\frac{d\phi}{dT}\Big)\right)&=&\frac{k\, e^{-2\phi/(d-1)}}{a_E^d} \\
g\left(e^{-2\phi/(d-1)}\Big(\frac{1}{a_E}\frac{da_E}{dT}+\frac{2}{d-1}\frac{d\phi}{dT}\Big)\right)&=&-\left(\frac{d}{a_E}\frac{da_E}{dT}+\frac{2}{d-1}\frac{d\phi}{dT}\right)^2e^{-\frac{4 \phi}{d-1}} 
\eea
We have written down the equations in terms of the Einstein frame scale factor and the physical dilaton. It is easy to see that setting the dilaton to a constant leads to a functional equations that determines $g(x)=-d^2 x^2$. Note that this is in tension with the supergravity expectation \eqref{g-expansion}. But if we ignore this and take \eqref{interesting} to be the definition of $f$, we can solve the system completely, and we find that the scale factor goes as a power law $a_E(T) \sim T^{1/d}$. 

The situation is essentially identical in string frame, because the two scale factors are related by a dilaton factor, which is constant in this ansatz. Note also that we could have worked very easily with the $a_E$ and $a$ variables as before, instead of working with the dilaton explicitly: holding the dilaton fixed is the same as holding the ratio of $a$ and $a_E$ fixed. 

\section{Concluding Comments}

There are two features that make the results of \cite{Hohm1, Hohm2} extremely interesting in our view:
\begin{itemize}
\item Firstly, their results provide all-order results in $\alpha'$ in a quite general setting. In particular, the result does not rely on supersymmetry or any of the usual assumptions that restrict dynamics. If anything, since the result relies only on the field content and the assumption of $O(d,d)$-duality, it is more general even than string theory. 
\item Most strikingly, their result applies to time-dependent cosmological settings, which is the one context where string theory has had enormous difficulties (possibly due to technical issues due to lack of control, but perhaps also due to lack of conceptual  clarity). An all order result in a cosmological setting, must clearly be of {\em some} use. 
\end{itemize}

But to take full advantage of these results, let us also put them in context, and state some challenges: 
\begin{itemize}
\item The results of \cite{Hohm1, Hohm2} are tree level in $g_s$. 
\item As emphasized in the introduction, the known (albeit controversial!) constructions of de Sitter vacua in string theory \cite{KKLT, LVS} require quantum corrections. It is unclear at the moment whether one can have tree level de Sitter vacua, even if one allows some non-perturbative singular sources. There have been many constructions of such tree level dS solutions in string theory, but they are all tachyonic. In fact this was one of the original motivations of \cite{Garg1, Garg2} for refining the initial dS swampland conjecture of \cite{Vafa1}. Therefore, an essential question in the construction of dS solutions in the present setting, is whether they are in fact stable. This has not been addressed at all, neither in \cite{Hohm1, Hohm2}, nor in our work here. Because of the presence of the unknown function $F(H)$, demonstrating (in)stability is likely to take some new ideas. 
\item It is worth emphasis that the dilaton is time-dependent in any candidate de Sitter solution, both in the string frame of \cite{Hohm1, Hohm2} as well as the Einstein frame here.
\item To construct all order corrections, one uses only duality invariance together with the massless field content, so it is in principle possible to do similar exercises for more general string theories with more fields. It will be very interesting to see if one can have dS solutions with NSNS/RR fluxes in some appropriate setting. 
\item More practically and less formally, is it possible to find solutions of the system that correspond to bouncing cosmologies or pre-Big Bang scenarios? The strategy that we have used in this paper, which is to start with the guess for the scale factor and then trying to see whether a consistent solution can be found, might be of some traction in this context as well. Interesting classes of string-inspired higher derivative theories (whose ultimate fate is yet to be settled)  with bouncing solutions, have appeared previously \cite{Siegel}. The higher derivative corrections there are chosen so that gravity becomes asymptotically free at short distances without introducing ghosts. It will be interesting to see if there is a connection between these two approaches, both seem to involve choosing a suitable function.
\end{itemize}

\section{Acknowledgments}

I thank the audience members at ICTS, Bangalore for a lively discussion during a String Seminar where the material in this note was presented. I also thank Mani C. Jha, K. V. Pavan Kumar, Ashoke Sen, P. N. Bala Subramanian and Zaid Zaz for discussions. 

\appendix


\end{document}